\documentclass[twocolumn,showpacs,preprintnumbers,prd,amssymb]{revtex4}
\usepackage{dcolumn}% Align table columns on decimal point
\newcommand{\be}{\begin{equation}}
\newcommand{\ee}{\end{equation}}
\newcommand{\ba}{\begin{eqnarray}}
\newcommand{\ea}{\end{eqnarray}}

\begin{document}
%\preprint{hep-th/00001}

\title{Angular Momentum and Energy--Momentum \\ Densities as Gauge Currents}
\author{M. Cal\c cada}
  \email{mcalcada@ift.unesp.br}
\author{J. G. Pereira}
 \email{jpereira@ift.unesp.br}
\affiliation{Instituto de F\'{\i}sica Te\'orica \\ Universidade Estadual Paulista\\
Rua Pamplona 145 \\
01405-900 S\~ao Paulo SP \\ Brazil}
\date{\today}

\begin{abstract}
If we replace the general {\em spacetime} group of diffeomorphisms by transformations taking
place in the {\em tangent space}, general relativity can be interpreted as a gauge theory, and in
particular as a gauge theory for the Lorentz group. In this context, it is shown that the angular
momentum and the energy--momentum tensors of a general matter field can be obtained from the
invariance of the corresponding action integral under transformations taking place, not in {\em
spacetime}, but in the {\em tangent space}, in which case they can be considered as gauge
currents.

\end{abstract}

\maketitle

%%%%%%%%%%%%%%%%%%%%%%
\section{Introduction}
%%%%%%%%%%%%%%%%%%%%%%

According to the Noether theorems~\cite{kopov}, energy--mo\-men\-tum conservation is related to
the invariance of the action integral under translations of the spacetime coordinates, and
angular--momentum conservation is related to the invariance of the action integral under Lorentz
transformations. As both translation and Lorentz transformations are perfectly well defined in the
Minkowski spacetime, the Noether theorems can be applied with no restrictions in this spacetime.
However, on a curved spacetime, neither translations nor Lorentz transformations can be defined in
a natural way~\cite{wald}. The problem then arises on how to define energy--momentum and
angular--momentum in the presence of gravitation, as in this case spacetime is represented by a
curved (pseudo) riemannian manifold.

In general relativity, the conservation of the energy--momentum tensor of any matter field is
usually obtained as a consequence of the invariance of the action integral in relation to the
spacetime group of diffeomorphisms (general coordinate transformations). Although this is usually
considered as acceptable for the energy--momentum tensor~\cite{trautman}, it leads to problems
when considering the angular momentum conservation, mainly in the case of spinor fields. In fact,
as the angular momentum conservation is related to the invariance of the action integral under
Lorentz transformations, and as there is no natural action of the full group of diffeomorphisms on
spinor fields~\cite{veltman}, the spin character of these fields has necessarily to be taken into
account by considering the action of the Lorentz group on the Minkowski tangent spacetime, where
its action is well defined. As a consequence, the Dirac equation in general relativity must
necessarily be written in terms of the spin connection, a connection assuming values in the
Lie algebra of the Lorentz group, and can never be written in terms of the spacetime Levi--Civita
(or Christoffel) connection.

In order to circumvent the above problems, let us then consider the following structure. At each
point of spacetime --- which in the presence of gravitation is a curved (pseudo) riemannian
manifold --- there is always a Minkowski tangent spacetime attached to it. Now, instead of
considering the spacetime group of diffeomorphism as the fundamental group behind gravitation, we
consider the {\em local} symmetry group of general relativity to be the Lorentz group~\cite{cp1},
whose action takes place in the tangent space. According to this construction, general relativity
can be reinterpreted as a gauge theory~\cite{hehl} for the Lorentz group~\cite{utiyama}, with the
indices related to the Minkowski space considered as {\em local} Lorentz indices in relation to
spacetime. As a consequence, the spin connection is to be considered as the fundamental field
representing gravitation. This means to use, instead of the Levi--Civita covariant derivative, the
Fock--Ivanenko operator, a covariant derivative that takes into account the spin content of the
field as defined in the Minkowski tangent space. This approach, mandatory for the case of
spinors~\cite{dirac}, can actually be used for any field, being in this sense more general than
the usual spacetime approach of general relativity.

By adopting the above described point of view, which means to reinterpret general relativity as
a gauge theory for the Lorentz group, the aim of the present paper will be to show that the
angular momentum and the energy--momentum tensors of a general matter field can be defined as
the Noether currents associated to the invariance of the action integral under local
transformations taking place not in spacetime, but in the tangent space. We begin in section II
with a review of the Lorentz transformation properties. In section III we introduce the gauge
potentials, define the Lorentz covariant derivative, and show how a very special tetrad field
naturally shows up in this formalism. This tetrad, as we are going to see, depends on the spin
connection, and this dependency will be crucial for obtaining the covariant conservation laws. In
section IV we obtain the gauge transformations of both the spin connection and the tetrad field. 
The roles played by the spin and the orbital parts of the Lorentz generators in these
transformations will also be analyzed. In section V we show how the angular momentum
conservation can be obtained as the Noether current associated to a transformation generated by
the spin generator of the Lorentz group. Then, by considering the Lorentz transformation of the
tetrad field, we show in section VI how the energy--momentum conservation can be obtained as the
Noether current associated to a transformation generated by the orbital generator of the Lorentz
group. Finally, in section VII, we comment on the results obtained.

%%%%%%%%%%%%%%%%%%%%%%%%%%%%%%%%%
\section{Lorentz Transformations}
%%%%%%%%%%%%%%%%%%%%%%%%%%%%%%%%%

We use the greek alphabet $\mu$, $\nu$, $\rho$,~$\dots=1,2,3,4$ to denote indices related to
spacetime, and the latin alphabet $a,b,c,\dots = 1,2,3,4$ to denote indices related to each one of
the Minkowski tangent spaces. The cartesian Minkowski coordinates, therefore, is denoted by
$\{x^a\}$, and its metric tensor is chosen to be
\be
\eta_{a b} = {\rm diag} (1, -1, -1, -1).
\ee
As is well known, the most general form of the generators of infinitesimal Lorentz transformations
is~\cite{ramond}
\be
J_{a b} = L_{a b} + S_{a b},
\label{fullrep}
\ee
where
\be
L_{a b} = i (x_a \partial_b - x_b \partial_a)
\label{loregen}
\ee
is the {\em orbital} part of the generators, and $S_{a b}$ is the {\em spin} part of the
generators, whose explicit form depends on the field under consideration. The generators $J_{a b}$
satisfy the commutation relation
\be
[J_{a b}, J_{c d}] = i \left(\eta_{b c} \, J_{a d} - \eta_{a c} \, J_{b d} -
\eta_{b d} \, J_{a c} + \eta_{a d} \, J_{b c} \right),
\label{commu}
\ee
which is to be identified with the Lie algebra of the Lorentz group. Each one of the generators
$L_{a b}$ and $S_{a b}$ satisfies the same commutation relation as $J_{a b}$, and commute with
each other.

A position dependent --- that is, local --- infinitesimal Lorentz
transformation is defined as
\be
\delta_L x^a = - \frac{i}{2} \, \epsilon^{c d} \, L_{c d} \, x^a,
\label{lore}
\ee
where $\epsilon^{c d} \equiv \epsilon^{c d}(x^\mu)$ are the transformation parameters. By using
the explicit form of $L_{c d}$, it becomes
\be
\delta_L x^a = - \epsilon^{a}{}_{d} \, x^d.
\label{lore1}
\ee
An interesting property of the Lorentz transformation of the Minkowski space coordinates is that
it is formally equivalent to a translation~\cite{kibble}. In fact, by using the explicit form
of $L_{c d}$, the transformation (\ref{lore}) can be rewritten in the form
\be
\delta_L x^a = - i \, \xi^c \, P_c \, x^a,
\label{relore}
\ee
which is a translation with
\be
\xi^c = \epsilon^c{}_d \, x^d
\label{constr}
\ee
as the transformation parameters, and
\be
P_c = - i \partial_c
\ee
as generators. In other words, an infinitesimal Lorentz transformation of the Minkowski
coordinates is equivalent to a translation with $\xi^c \equiv \epsilon^c{}_d \, x^d$ as the
parameters. Actually, this is a property of the Lorentz generators $L_{a b}$, whose action can
always be reinterpreted as a translation. The reason for such equivalence is that, because the
Minkowski spacetime is transitive under translations, every two points related by a Lorentz
transformation can also be related by a translation. Notice that the inverse is not true.

Let us consider now a general matter field $\Psi(x^\mu)$, which is function of the spacetime
coordinates $\{x^\mu\}$. Under an infinitesimal {\em local} Lorentz transformation of the
tangent--space coordinates, the field $\Psi$ will change according to~\cite{ramond}
\be
\delta_J \Psi \equiv \Psi^\prime(x) - \Psi(x) =
- \frac{i}{2} \, \epsilon^{a b} J_{a b} \Psi(x).
\label{lt1}
\ee
The explicit form of the orbital generators $L_{a b}$, given by (\ref{loregen}), is the same for
all fields, whereas the explicit form of the spin generators $S_{a b}$ depends on the spin
of the field $\Psi$. Notice furthermore that the orbital generators $L_{a b}$ are able to
act in the spacetime argument of $\Psi(x^\mu)$ due to the relation
\[
\partial_a = (\partial_a x^\mu) \, \partial_\mu.
\]
By using the explicit form of $L_{a b}$, the Lorentz transformation (\ref{lt1}) can be
rewritten as
\be
\delta_J \Psi = - \epsilon^{a b} \, x_b \, \partial_a \Psi -
\frac{i}{2} \, \epsilon^{a b}{} S_{a b} \Psi,
\label{lt2}
\ee
or equivalently,
\be
\delta_J \Psi = - i \xi^c P_c \Psi -
\frac{i}{2} \, \epsilon^{a b} S_{a b} \Psi,
\label{lt3}
\ee
where use has been made of Eq.~(\ref{constr}). In other words, the {\em orbital} part of the
transformation can be reduced to a translation, and consequently the Lorentz transformation of
a general field $\Psi$ can be rewritten as a {\em translation} plus a strictly {\em spin} Lorentz
transformation. Notice however that, as
\be
[ P_c, S_{a b} ] = 0,
\label{ps0}
\ee
the transformation (\ref{lt3}) is not a Poincar\'e, but a Lorentz transformation.

As a final remark, it is important to notice that, instead of four scalar functions, the
coordinates $x^a$ of the Minkowski spacetime can also be interpreted as a vector field
$x^a(x^\mu)$. In this case, however, the Lorentz generators must be written in the vector
representation
\be
\left( S_{cd}\right)^a{}_b = i \left( \delta_c{}^a \, \eta_{d b} -
\delta_d{}^a \, \eta_{c b} \right).
\label{spinge}
\ee
Consequently, its Lorentz transformation will be written as
\be
\delta_S x^a = - \frac{i}{2} \, \epsilon^{c d} \, \left( S_{cd}\right)^a{}_b \, x^b,
\label{slore}
\ee
which yields
\be
\delta_S x^a = \epsilon^{a}{}_{d} \, x^d.
\label{lore2}
\ee
Therefore, from Eqs.~(\ref{lore1}) and (\ref{lore2}) we see that a Lorentz transformation of the
Minkowski coordinates written with the complete generator $J_{c d}$ vanishes identically:
\be
\delta_J x^a \equiv - \frac{i}{2} \, \epsilon^{c d} \, J_{cd} \, x^a = 0.
\label{lore3}
\ee
The interpretation of this result is that, under a Lorentz transformation generated by $J_{c d}$,
all vector fields $V^a(x)$ undergo a transformation at the same point:
\[
\delta_J V^a \equiv V^{a \prime}(x) - V^a(x) =
- \frac{i}{2} \, \epsilon^{c d} J_{c d} \, V^a.
\]
In the specific case of the coordinate itself, which is also a Lorentz vector field, the
transformations generated by $S_{c d}$ and $L_{c d}$ cancel each other, yielding a vanishing net
result.

%%%%%%%%%%%%%%%%%%%%%%%%%%%%%%%%%%%%%%
\section{Lorentz Covariant Derivative}
%%%%%%%%%%%%%%%%%%%%%%%%%%%%%%%%%%%%%%

In a gauge theory for the Lorentz group, the fundamental field representing gravitation is the
spin connection ${\stackrel{~~\circ}{\mathcal A}}{}_\mu$, a field assuming values in the Lie
algebra of the Lorentz group,
\be
{\stackrel{~~\circ}{\mathcal A}}{}_\mu =
\frac{1}{2} {\stackrel{~\circ}{A}}{}^{a b}{}_\mu \, J_{a b}.
\ee
Equivalently, we can write
\be
{\stackrel{~~\circ}{\mathcal A}}{}_\mu = c^{-2} B^a{}_\mu \, P_a +
\frac{1}{2} {\stackrel{~\circ}{A}}{}^{a b}{}_\mu \, S_{a b},
\ee
where a new gauge potential $B^a{}_\mu$  assuming values in the Lie algebra of
the translation group, has been defined~\cite{cp1}
\be
c^{-2} \, B^a{}_\mu = {\stackrel{~\circ}{A}}{}^{a}{}_{b \mu} \, x^b,
\label{crucial}
\ee
with the velocity of light $c$ introduced for dimensional reasons. It is important to remark once
more that, despite the existence of a gauge field related to translations, and another one related
to the Lorentz group, the structure group underlying this construction is not the Poincar\'e, but
the Lorentz group.

We consider now the Lorentz covariant derivative of the matter field $\Psi$, whose general form
is~\cite{livro} 
\be
{\stackrel{\circ}{\mathcal D}}{}_c \Psi = \partial_c \Psi +
\frac{1}{2} \, {\stackrel{~\circ}{A}}{}^{a b}{}_c \,
\frac{\delta_J \Psi}{\delta \epsilon^{a b}}.
\ee
Substituting the transformation (\ref{lt2}), it becomes~\cite{cp1}
\be
{\stackrel{\circ}{\mathcal D}}{}_c \Psi =
h^\mu{}_c \, {\stackrel{\circ}{\mathcal D}}{}_\mu \Psi,
\label{locoder}
\ee
where $h^\mu{}_c$ is the inverse of the tetrad field
\be
h^c{}_\mu = \partial_\mu x^c + {\stackrel{~\circ}{A}}{}^c{}_{d \mu} x^d \equiv
\partial_\mu x^c + c^{-2} B^c{}_\mu,
\label{tetrada}
\ee
and
\be
{\stackrel{\circ}{\mathcal D}}{}_\mu = \partial_\mu -
\frac{i}{2} \, {\stackrel{~\circ}{A}}{}^{a b}{}_\mu \, S_{a b}
\label{fi}
\ee
is the Fock--Ivanenko covariant derivative operator~\cite{fi}. According to this construction,
therefore, the {\em orbital} part of the Lorentz generators is reduced to a translation, which
gives then rise to a tetrad that depends on the spin connection. Because its action reduces
ultimately to a translation, the {\em orbital} generator $L_{a b}$ is the responsible for the
universality of gravitation. In fact, as $L_{a b}$ acts in the fields through their arguments, all
fields will respond equally to its action. Notice also that, whereas the tangent space indices are
raised and lowered with the metric $\eta_{a b}$, spacetime indices are raised and lowered with the
riemannian metric
\be
g_{\mu \nu} = h^a{}_\mu \, h^b{}_\nu \, \eta_{a b}.
\label{metric}
\ee

It is important to remark that the Fock--Ivanenko derivative has the definition
\[
{\stackrel{\circ}{\mathcal D}}{}_c \Psi = \partial_c \Psi +
\frac{1}{2} \, {\stackrel{~\circ}{A}}{}^{a b}{}_c \,
\frac{\delta_S \Psi}{\delta \epsilon^{a b}},
\]
where~\cite{ramond}
\be
\delta_S \Psi \equiv \Psi^\prime(x^\prime) - \Psi(x) =
- \frac{i}{2} \, \epsilon^{a b} S_{a b} \Psi.
\label{tlt1}
\ee
Accordingly, the Fock-Ivanenko covariant derivative of the vector field $x^c(x^\mu)$ is 
\be
{\stackrel{\circ}{\mathcal D}}{}_\mu x^c = \partial_\mu x^c +
\frac{1}{2} {\stackrel{~\circ}{A}}{}^{a b}{}_\mu \,
\frac{\delta_S x^c}{\delta \epsilon^{a b}}.
\ee
Using the transformation (\ref{lore2}), it becomes
\be
{\stackrel{\circ}{\mathcal D}}{}_\mu x^c = \partial_\mu x^c +
{\stackrel{~\circ}{A}}{}^c{}_{b \mu} x^b \equiv h^c{}_\mu,
\ee
which shows that the tetrad coincides with the Fock-Ivanenko covariant derivative of the
vector field $x^c(x^\mu)$.

%%%%%%%%%%%%%%%%%%%%%%%%%%%%%%%
\section{Gauge Transformations}
%%%%%%%%%%%%%%%%%%%%%%%%%%%%%%%

Under a local Lorentz transformation generated by
\be
U = \exp  \left[ - \frac{i}{2} \, \epsilon^{a b} \, S_{a b} \right],
\ee
the covariant derivative ${\stackrel{\circ}{\mathcal D}}{}_a \Psi$ will change according to
\be
{\stackrel{\circ}{\mathcal D}}{}^\prime_a \Psi^\prime(x^\prime) =
U {\stackrel{\circ}{\mathcal D}}{}_a \Psi(x).
\label{detrans}
\ee
As
\be
{\stackrel{\circ}{\mathcal D}}{}_a \Psi(x) =
h^\mu{}_a \, {\stackrel{\circ}{\mathcal D}}{}_\mu \Psi(x),
\ee
and taking into account that $h^{\prime \mu}{}_a(x^\prime) = U_a{}^b \; h^\mu{}_b(x)$, with
$U_a{}^b$ the usual element of the Lorentz group in the vector representation, we can rewrite
Eq.~(\ref{detrans}) in the form
\be
h^{\prime \mu}{}_a(x^\prime) \; {\stackrel{\circ}{\mathcal D}}{}^\prime_\mu \Psi^\prime(x^\prime) =
U_a{}^b \, h^\mu{}_b(x) \; U {\stackrel{\circ}{\mathcal D}}{}_\mu \Psi(x).
\ee
It then follows that
\be
{\stackrel{\circ}{\mathcal D}}{}^\prime_\mu \Psi^\prime(x^\prime) = 
U {\stackrel{\circ}{\mathcal D}}{}_\mu \Psi(x),
\ee
or equivalently,
\be
{\stackrel{\circ}{\mathcal D}}{}^\prime_\mu = U {\stackrel{\circ}{\mathcal D}}{}_\mu U^{-1}.
\ee
Using the Fock--Ivanenko derivative (\ref{fi}), we obtain the usual gauge transformation
\be
{\stackrel{~\circ}{A}}{}^\prime_\mu = U {\stackrel{~\circ}{A}}{}_\mu U^{-1} +
i U \partial_\mu U^{-1}.
\label{trans1}
\ee
The infinitesimal form of $U$ is
\be
U \simeq 1 - \frac{i}{2} \, \epsilon^{c d} \, S_{c d}.
\label{trans2}
\ee
By using the commutation relation (\ref{commu}) for $S_{a b}$, we get from (\ref{trans1})
\be
\delta_S {\stackrel{~\circ}{A}}{}^{c d}{}_\mu = - \left(\partial_\mu \epsilon^{c d} +
{\stackrel{~\circ}{A}}{}^c{}_{a \mu} \, \epsilon^{a d} +
{\stackrel{~\circ}{A}}{}^d{}_{a \mu} \, \epsilon^{c a} \right) \equiv
- {\stackrel{\circ}{\mathcal D}}{}_\mu  \epsilon^{c d}.
\label{atrans}
\ee
Notice that, as ${\stackrel{~\circ}{A}}{}^{c d}{}_\mu$ does not respond to the orbital generators,
we have that $\delta_S {\stackrel{~\circ}{A}}{}^{c d}{}_\mu$ = $\delta_J
{\stackrel{~\circ}{A}}{}^{c d}{}_\mu$.

Let us obtain now the infinitesimal Lorentz transformations of the tetrad
field. First of all, we have the transformation generated by $S_{a b}$, which yields the {\em
total} change in the tetrad, that is, $\delta_S h^a{}_\mu \equiv h^{\prime a}{}_\mu(x^\prime) -
h^a{}_\mu(x)$. From (\ref{tetrada}) we see that
\be
\delta_S h^a{}_\mu = \partial_\mu (\delta_S x^a) + (\delta_S {\stackrel{~\circ}{A}}{}^a{}_{d \mu})
\, x^d + {\stackrel{~\circ}{A}}{}^a{}_{d \mu} \, (\delta_S x^d).
\label{total}
\ee
Using the transformations (\ref{lore2}) and (\ref{atrans}), we get
\be
\delta_S h^a{}_\mu = \epsilon^a{}_c \,  h^c{}_\mu \equiv
- \frac{i}{2} \, \epsilon^{c d} \, \left( S_{cd}\right)^a{}_b \, h^b{}_\mu,
\label{usualt}
\ee
as it should be since $h^a{}_\mu$ is a Lorentz vector field in the tangent--space index. On
the other hand, the tetrad transformation generated by $J_{a b}$ corresponds to a transformation
at the same $x^a$, that is, $\delta_J h^a{}_\mu \equiv h^{a \prime}{}_\mu(x) - h^a{}_\mu(x)$. Such
a transformation can be obtained from (\ref{tetrada}) by keeping $x^a$ fixed, and substituting
$\delta_J {\stackrel{~\circ}{A}}{}^a{}_{b \mu} \equiv \delta_S {\stackrel{~\circ}{A}}{}^a{}_{b
\mu}$ as given by Eq.~(\ref{atrans}). The result is
\be
\delta_J h^a{}_\mu = - x_b \, {\stackrel{\circ}{\mathcal D}}{}_\mu \epsilon^{a b}.
\label{htrans}
\ee
Finally, there is also the transformation generated by the orbital generator $L_{a b}$. As the
spin connection transformation is generated by the spin generators $S_{a b}$ only --- see
Eqs.~(\ref{trans1}) and (\ref{trans2}) --- this corresponds to a transformation due to the
variation of the coordinate $x^a$ only, with ${\stackrel{~\circ}{A}}{}^a{}_{d \mu}$ fixed:
$\delta_L h^a{}_\mu
\equiv h^a{}_\mu(x^\prime) - h^a{}_\mu(x)$.  From Eq.~(\ref{tetrada}), we see that such a
transformation is given by
\be
\delta_L h^a{}_\mu =
\partial_\mu (\delta_L x^a) + {\stackrel{~\circ}{A}}{}^a{}_{d \mu} \, (\delta_L x^d).
\label{dx}
\ee
Substituting (\ref{lore1}), and making use of the definition (\ref{constr}), we get
\be
\delta_L h^a{}_\mu  = - {\stackrel{\circ}{\mathcal D}}{}_\mu \xi^a.
\label{xtrans}
\ee
This transformation shows that the tetrad behaves as a translational gauge potential under a
Lorentz transformation of the tangent space coordinates, in which only the change due to the
variation of the coordinates is considered. In other words, the tetrad behaves like a
translational gauge potential under a Lorentz transformation generated by the orbital generator
$L_{a b}$, whose action, as we have already seen, can always be reinterpreted as a translation.

Notice finally that, by using the above results, the transformation (\ref{usualt}) can be
rewritten in the form
\be
\delta_S h^a{}_\mu \equiv \delta_J h^a{}_\mu - \delta_L h^a{}_\mu =
- x^b \, {\stackrel{\circ}{\mathcal D}}{}_\mu \epsilon^{a b} +
{\stackrel{\circ}{\mathcal D}}{}_\mu \xi^a.
\label{star}
\ee
We remark that this result is easily seen to be equivalent to (\ref{usualt}) by using the fact
that the tetrad is the covariant derivative of the tangent space coordinate $x^a$.

%%%%%%%%%%%%%%%%%%%%%%%%%%%%%%%%%%%%%%%
\section{Angular Momentum Conservation}
%%%%%%%%%%%%%%%%%%%%%%%%%%%%%%%%%%%%%%%

Let us consider now a general matter field $\Psi$ with the action integral
\be
S = \frac{1}{c} \int {\mathcal L} \; d^4x \equiv
\frac{1}{c} \int L \, h \ d^4x,
\label{action}
\ee
where $h = \det(h^a{}_\mu) = \sqrt{-g}$, with $g = \det(g_{\mu \nu})$. We assume a first-order
formalism, according to which the lagrangian depends only on the fields and on their first
derivatives. Under a local Lorentz transformation of the tangent--space coordinates, both
${\stackrel{~\circ}{A}}{}^a{}_{b \mu}$ and $h^a{}_\mu$ will change. The transformation of the spin
connection ${\stackrel{~\circ}{A}}{}^a{}_{b \mu}$ is generated by the spin part of the Lorentz
generators, whereas the transformation of the tetrad $h^a{}_\mu$ is generated by both the spin and
the orbital parts.

Let us consider first the response of the action integral due to the change of the
Lorentz gauge potential ${\stackrel{~\circ}{A}}{}^{ab}{}_\mu$. As a Lorentz scalar, the action
integral is invariant under a local Lorentz transformation generated by $S_{a b}$. Under such
a transformation, it changes according to  
\begin{equation}
     \delta S = \frac{1}{2 c} \int \left[\frac{\partial {\mathcal L}}
    {\partial {\stackrel{~\circ}{A}}{}^{ab}{}_\mu} - \partial_\rho \, \frac{\partial {\mathcal L}}
    {\partial \partial_\rho {\stackrel{~\circ}{A}}{}^{ab}{}_\mu} \right] \;
\delta_S {\stackrel{~\circ}{A}}{}^{ab}{}_\mu \; d^4x,
\end{equation}
where we have not written the variation in relation to the field $\Psi$ 
because it gives the associated field equation~\cite{landau}. Introducing the 
notation
\begin{equation}
   \left[\frac{\partial {\mathcal L}}
   {\partial {\stackrel{~\circ}{A}}{}^{ab}{}_\mu} - \partial_\rho \, \frac{\partial {\mathcal L}}
   {\partial \partial_\rho {\stackrel{~\circ}{A}}{}^{ab}{}_\mu} \right] \equiv  \, \frac{\delta 
   {\mathcal L}}{\delta {\stackrel{~\circ}{A}}{}^{ab}{}_\mu} =  h \, {\mathcal J}^\mu{}_{ab},
\label{amt}
\end{equation}
where ${\mathcal J}^\mu{}_{ab}$ is the angular momentum tensor, it follows that 
\begin{equation}
     \delta S = \frac{1}{2 c} \int {\mathcal J}^\mu{}_{ab} \; \delta_S
{\stackrel{~\circ}{A}}{}^{ab}{}_\mu \; h \; d^4x.
\label{av}
\end{equation}
Substituting the transformation (\ref{atrans}), integrating by parts, and  neglecting the surface
term, we obtain
\be
\delta S = - \frac{1}{2 c} \int {\stackrel{\circ}{\mathcal D}}{}_\mu
(h {\mathcal J}^\mu{}_{ab})  \; \epsilon^{ab}  \; d^4x.
\ee
Due to the arbitrariness of $\epsilon^{ab}$, it follows from the invariance of the action integral
under local Lorentz transformations that
\be
{\stackrel{\circ}{\mathcal D}}{}_\mu (h {\mathcal J}^\mu{}_{ab}) = 0.
\ee
Using the identity
\be 
\partial_\mu h = h \; {\stackrel{\circ}{\Gamma}}{}^\lambda{}_{\lambda \mu},
\label{identidade}
\ee
with ${\stackrel{\circ}{\Gamma}}{}^\lambda{}_{\lambda \mu} =
{\stackrel{\circ}{\Gamma}}{}^\lambda{}_{\mu \lambda}$ the Levi--Civita connection of the metric
(\ref{metric}), we get
\begin{equation}
\partial_\mu {\mathcal J}^\mu{}_{a b} + {\stackrel{\circ}{\Gamma}}{}^\mu{}_{\lambda 
\mu} \, {\mathcal J}^\lambda{}_{ab} - {\stackrel{~\circ}{A}}{}^c{}_{a \mu} \, 
{\mathcal J}^\mu{}_{c b} - {\stackrel{~\circ}{A}}{}^c{}_{b \mu} \, {\mathcal J}^\mu{}_{a c} = 0,
\end{equation}
which is the usual covariant conservation law of the angular momentum tensor in general relativity.
According to this construction, therefore, we see that the angular momentum conservation is
related to the response of the action integral under a Lorentz transformation of the spin
connection, which is a transformation generated by the spin generator $S_{a b}$.

%%%%%%%%%%%%%%%%%%%%%%%%%%%%%%%%%%%%%%%
\section{Energy--Momentum Conservation}
%%%%%%%%%%%%%%%%%%%%%%%%%%%%%%%%%%%%%%%

The angular momentum tensor can be rewritten in the form
\be
{\mathcal J}^\mu{}_{a b} = - {\mathcal T}^\rho{}_c \;
\frac{\delta h^c{}_\rho}{\delta {\stackrel{~\circ}{A}}{}^{a b}{}_\mu},
\label{inter}
\ee
where
\be
{\mathcal T}^\rho{}_c = - \frac{1}{h} \frac{\delta {\mathcal L}}{\delta h^c{}_\rho} \equiv
\frac{\partial {\mathcal L}}{\partial h^c{}_\rho} -
\partial_\lambda \, \frac{\partial {\mathcal L}}{\partial \partial_\lambda h^c{}_\rho}
\label{tem}
\ee
is the energy-momentum tensor. From Eq.~(\ref{tetrada}), we see that
\[
\frac{\delta h^c{}_\rho}{\delta {\stackrel{~\circ}{A}}{}^{a b}{}_\mu} = \delta^\mu{}_\rho
\left( \delta^c{}_a \, x_b - \delta^c{}_b \, x_a \right).
\]
Therefore, Eq.~(\ref{inter}) becomes
\be
{\mathcal J}^\mu{}_{a b} = x_a \, {\mathcal T}^\mu{}_b - x_b \, {\mathcal T}^\mu{}_a,
\label{xtxt}
\ee
which is the usual expression of the {\em total} angular momentum tensor in terms of the symmetric
energy-momentum tensor~\cite{angu}. Reversing the argument, we can say that the usual relation
between ${\mathcal J}^\mu{}_{a b}$ and ${\mathcal T}^\mu{}_a$ requires a tetrad of the form
(\ref{tetrada}).

Substituting now Eq.~(\ref{inter}) in the transformation (\ref{av}), it follows that
\be
\delta S = - \frac{1}{2c} \int {\mathcal T}^\rho{}_c \;
\frac{\delta h^c{}_\rho}{\delta {\stackrel{~\circ}{A}}{}^{a b}{}_\mu} \;
\delta_S {\stackrel{~\circ}{A}}{}^{a b}{}_\mu \; h \; d^4x,
\label{deltas1}
\ee
or equivalently,
\be
\delta S = - \frac{1}{c} \int {\mathcal T}^\rho{}_c \;
\delta_S h^c{}_\rho \; h \; d^4x.
\label{deltas2}
\ee
Substituting $\delta_S h^c{}_\rho$ as given by Eq.~(\ref{star}), integrating both terms by parts
and neglecting the corresponding surface terms, we obtain
\be
\delta S = \frac{1}{c} \int \left[{\stackrel{\circ}{\mathcal D}}{}_\mu
(h {\mathcal J}^\mu{}_{ab}) \; \epsilon^{ab} -
{\stackrel{\circ}{\mathcal D}}{}_\mu (h {\mathcal T}^\mu{}_a) \; \xi^a \right] \; d^4x.
\ee
Using the fact that the angular momentum is covariantly conserved, we get
\be
\delta S = - \frac{1}{c} \int
{\stackrel{\circ}{\mathcal D}}{}_\mu (h {\mathcal T}^\mu{}_a) \; \xi^a \; d^4x.
\label{ele}
\ee
Due to the arbitrariness of $\xi^a$, it follows from the invariance of the action integral under
a local Lorentz transformation that
\be
{\stackrel{\circ}{\mathcal D}}{}_\mu (h {\mathcal T}^\mu{}_a) = 0.
\ee
Using the identity (\ref{identidade}), this expression can be rewritten in the form
\be
\partial_\mu {\mathcal T}^\mu{}_a +
{\stackrel{\circ}{\Gamma}}{}^\mu{}_{\lambda \mu} \, {\mathcal T}^\lambda{}_a -
{\stackrel{~\circ}{A}}{}^c{}_{a \mu} \, {\mathcal T}^\mu{}_c = 0,
\ee
which is the usual covariant conservation law of general relativity.

It is important to notice that the energy--momentum covariant conservation in this case turns out
to be related to the response of the action integral under a transformation of the tetrad field
generated by the orbital generators $L_{ab}$, which as we have already seen are transformations
that can be reinterpreted as translations. In fact, after integrating (back) by parts and
neglecting the surface term, Eq.~(\ref{ele}) can be rewritten in the form
\be
\delta S = \frac{1}{c} \int {\mathcal T}^\mu{}_a \; \delta_L h^a{}_\mu \; h \; d^4x,
\label{ela}
\ee
which holds provided the angular momentum ${\mathcal J}^\mu{}_{ab}$ is covariantly conserved.
Furthermore, it is easy to see that $\delta_L h^a{}_\mu$, given by Eq.~(\ref{xtrans}), induces in
the metric tensor (\ref{metric}) the transformation
\be
\delta_L g_{\mu \nu} = - {\stackrel{\circ}{\nabla}}{}_\mu \xi_\nu -
{\stackrel{\circ}{\nabla}}{}_\nu \xi_\mu,
\label{lie}
\ee
where $\xi_\mu = \xi_a \, h^a{}_\mu$, and ${\stackrel{\circ}{\nabla}}{}_\mu$ is the
Levi-Civita covariant derivative. As is well known, this equation represents the response of
$g_{\mu \nu}$ to a general transformation of the spacetime coordinates, and its use in the
Noether theorem yields the covariant conservation law of the matter energy--momentum tensor in
the usual context of general relativity~\cite{landau}.

%%%%%%%%%%%%%%%%%%%%%%%
\section{Final Remarks}
%%%%%%%%%%%%%%%%%%%%%%%

According to the Noether theorems, energy--mo\-men\-tum conservation is related to the invariance
of the action integral under spacetime translations, and angular momentum conservation is related
to the invariance of the action integral under spacetime Lorentz transformations. However, as is
well known, in the presence of gravitation spacetime becomes a (pseudo) riemannian manifold. As
the above transformations cannot be defined on such spacetimes~\cite{wald}, it is necessary to
introduce a local procedure in which the corresponding covariant conservation laws can be obtained
from the invariance of the action integral under transformations taking place in the Minkowski
tangent space, where they are well defined.

By considering general relativity as a gauge theory for the Lorentz group, where the spin
connection --- that is, the Lorentz gauge potential --- is the fundamental field representing
gravitation, we have shown that it is possible to obtain the angular momentum and the
energy--momentum covariant conservation laws from the invariance of the action integral under
transformations taking place in the tangent space. The crucial point of this formalism is the
Lorentz covariant derivative (\ref{locoder}), in which the action of the orbital Lorentz
generators reduces to a translation, giving then rise to a {\em translational} gauge potential
$c^{-2} B^a{}_\mu = {\stackrel{~\circ}{A}}{}^a{}_{b \mu} x^b$ that appears as the nontrivial part
of the tetrad field:
\be
h^a{}_\mu = \partial_\mu x^a + {\stackrel{~\circ}{A}}{}^a{}_{b \mu} x^b.
\label{again}
\ee
We remark that this constraint between $h^a{}_\mu$ and ${\stackrel{~\circ}{A}}{}^a{}_{b \mu}$
yields naturally the usual relation, given by Eq.~(\ref{xtxt}), between the energy--momentum and
the angular momentum tensors, showing in this way the consistency of the tetrad (\ref{again}). In
this approach, the covariant conservation law of the angular momentum tensor turns out to be
related to the response of the action integral under Lorentz transformations of the spin
connection ${\stackrel{~\circ}{A}}{}^a{}_{b \mu}$, which is a transformation generated by the spin
part of the Lorentz generators. On the other hand, the energy--momentum conservation turns out to
related to the response of the action integral under a Lorentz transformation of the tetrad field.
Differently from ${\stackrel{~\circ}{A}}{}^a{}_{b \mu}$, the tetrad field $h^a{}_\mu$ responds
simultaneously to both the spin and the orbital Lorentz generators. The part related to spin
generator $S_{a b}$ yields again the conservation of the angular momentum tensor, written now in
the form (\ref{xtxt}). The part related to the orbital generator $L_{a b}$ yields the conservation
of the energy--momentum tensor, a result consistent with the fact that the Lorentz transformation
generated by $L_{a b}$ can always be reduced to a translation. In fact, the tetrad transformation
generated by $L_{a b}$, given by Eq.~(\ref{xtrans}), induces in the metric tensor $g_{\mu \nu}$
the transformation (\ref{lie}), which is the usual transformation of $g_{\mu \nu}$ under a general
transformation of the spacetime coordinates, and which yields the covariant conservation law of
the matter energy--momentum tensor in the usual context of general relativity. We have in this way
established a relation between spacetime diffeomorphisms and tangent space Lorentz transformations
generated by the orbital generator $L_{a b}$. This is a crucial result in the sense that it is the
responsible for obtaining the covariant conservation law for the energy--momentum tensor under
transformations taking place in the tangent space. We notice in passing that even in the tetrad
approach to general relativity, as the tetrad is {\em invariant} under a {\em true translation} of
the tangent space coordinates, no energy--momentum covariant conservation law can be obtained.
Summing up, with this construction we have succeeded in obtaining an internal Noether theorem from
which the covariant conservation laws for angular momentum and energy--momentum tensors are
obtained from the invariance of the action integral under ``internal'' --- that is, tangent space
--- transformations. Accordingly, the associated densities can be considered as ``gauge'' currents.

\begin{acknowledgments}
The authors thank Y. Obukhov for enlightening discussions. They also thank CNPq-Brazil and
FAPESP-Brazil for financial support.
\end{acknowledgments}

\end{document}